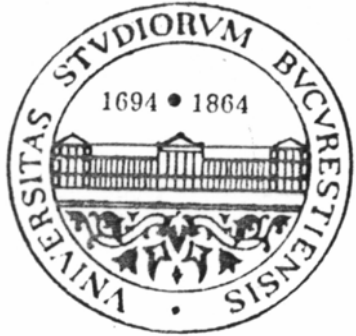

# BUCHAREST UNIVERSITY

## EXPERIMENTAL PARTICLE PHYSICS GROUP

# Lindhard Factors and Concentrations of Primary Defects in Semiconductor Materials for uses in HEP

I. Lazanu[a1] and S. Lazanu[b2]

[a] University of Bucharest, Department of Nuclear Physics,
P.O. Box MG-11, Bucharest-Magurele, Romania

[b] National Institute for Materials Physics, P.O.Box MG-7, Bucharest-Magurele, Romania

**Abstract**

**This report is a compilation of the authors' calculations of the Lindhard factors, and of the concentration of the primary radiation induced defects per unit fluence in different materials for HEP uses.**

---

[1] e-mail: i_lazanu@yahoo.co.uk
[2] e-mail: lazanu@alpha1.infim.ro

Recent developments of accelerator machines conduce to the necessity of the study of radiation effects in a multitude of materials, in intense fields of a large variety of particles, in a high range of energy and at different fluxes. A special interest is represented by the evaluation of these effects in silicon or in other possible candidates for uses as detectors or different devices in these environments.

This report is a compilation of the authors' calculations of the Lindhard factors, and of the concentration of the primary radiation induced defects per unit fluence in different materials for HEP uses.

# I. Lindhard factors

An essential factor in the evaluation of radiation effects is the knowledge of the energy partition of slowing down particles, between ionisation and displacements, in the lattice of the target. The result of the primary interaction of the incident particle with the target nuclei are ions of charge and mass numbers smaller or equal to the corresponding ones of the atoms of the medium. Lindhard theory gives a detailed description of the calculation procedure for the cases where the slowing down particles (primary particles or recoil nuclei resulting from interactions between primary beam and the semiconductor) are identical to the medium ones, and for monoelement materials, in particular for silicon.

The curves presented in this report are the Lindhard factors for diamond, SiC and $A^3B^5$ compounds (GaP, GaAs, InP, InAs, and InSb), calculated in the frame of the Lindhard hypothesis [*J. Lindhard, V. Nielsen, M. Schraff, P. V. Thomsen, Mat. Phys. Medd. Dan. Vid. Sesk. 33 (1963) 1*] and in the analytical approximations developed in the authors' paper [*S. Lazanu, I. Lazanu, Nucl. Instr. And Meth. A 462 (2001) 530*] for the cases when particles do not belong to the medium, both for monoelement semiconductors and compound media. For silicon, the Lindhard factors were calculated and published by Simon and the co-workers [*G. W. Simon, J. M. Denney, R. G. Downing, Phys. Rev. 129 (1963) 2545*] and are not included in this compilation.

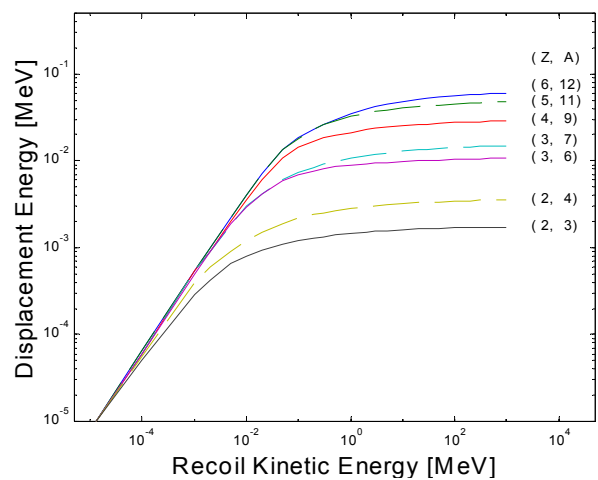

Figure 1. Displacement energy versus recoil energy in diamond - from reference [*I. Lazanu et al., Nucl. Instr. Meth. Phys. Res. A 406, 259 (1998)*].

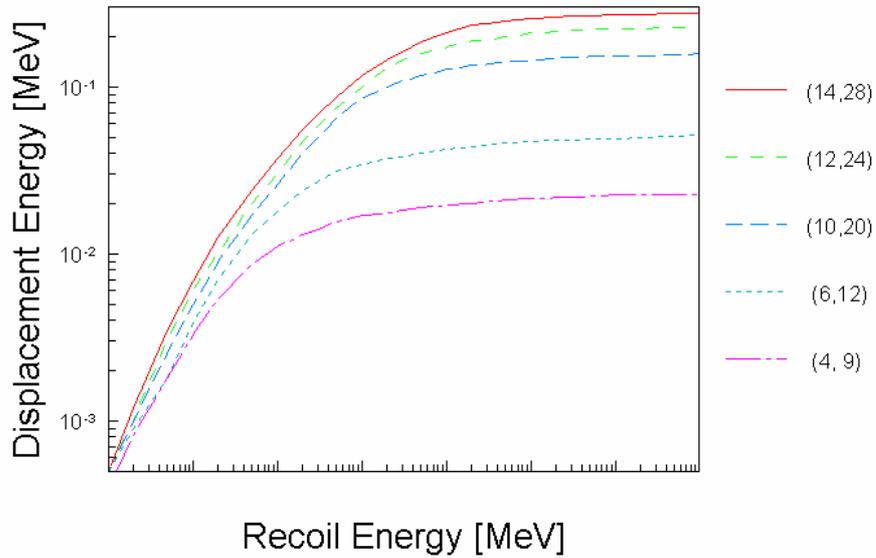

Figure 2. Dependence of the energy channelled in displacements as a function of the energy of the ion in the lattice of SiC. The curves are from reference [*S. Lazanu et. al., Nucl. Instr. Meth. Phys. Res. A 485, 768 (202)*].

The same type of dependencies for $A^3B^5$ compounds are represented in the figures that follows: in Figure 3 for GaAs, in Figure 4 for GaP, in Figure 5 for InAs, in Figure 6 for InP, and in Figure 7 for InSb. For compounds with remote elements in the periodic table, as is the case of, e.g. GaP and InP, the families of ions coming from the two elements are easily identifiable. The curves are from reference [*S. Lazanu et al., Nucl. Instr. Meth. Phys. Res. A 413, 242 (1998)*][1].

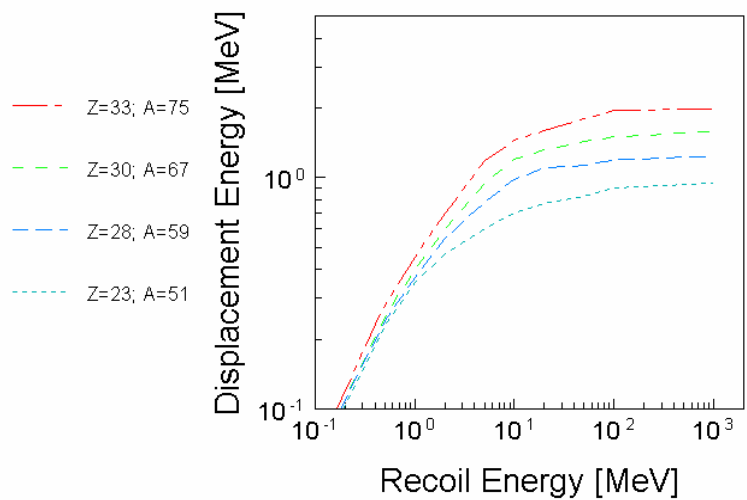

Figure 3. Lindhard factors in GaAs

---

[1] The scale error in displacement energies from the original paper was corrected.

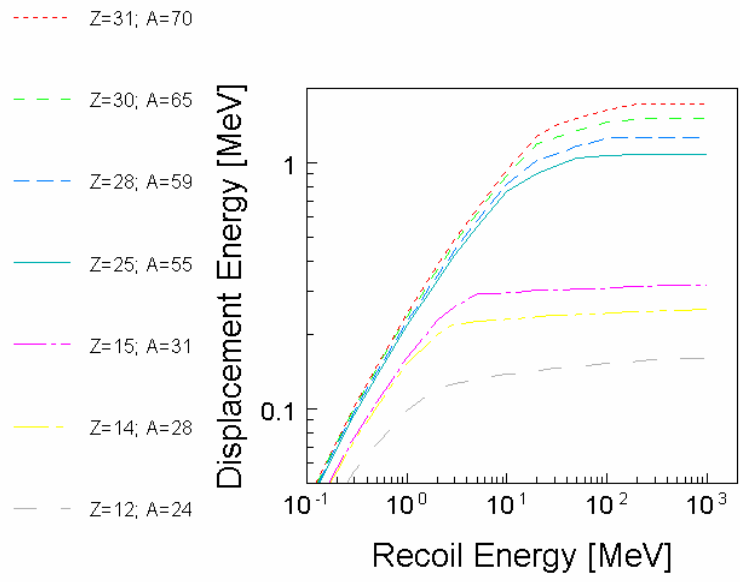

Figure 4. Lindhard factors in GaP

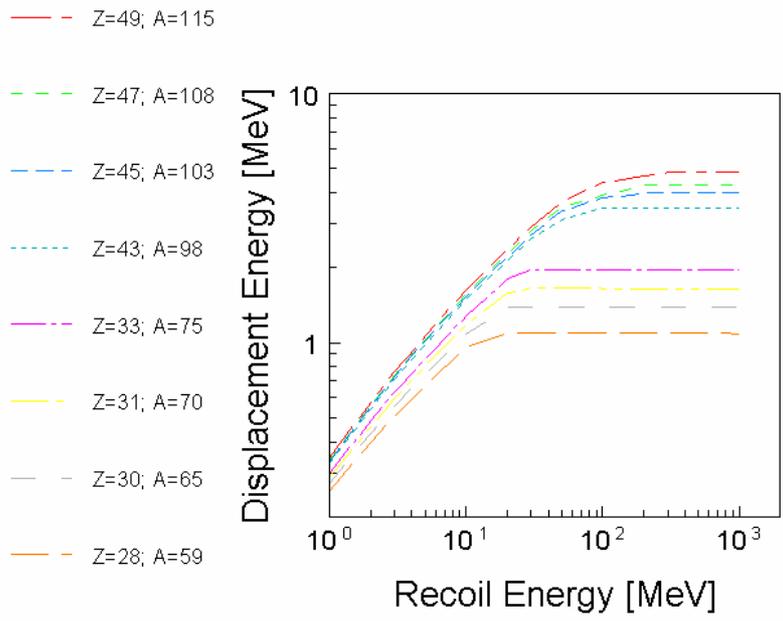

Figure 5. Lindhard factors in InAs

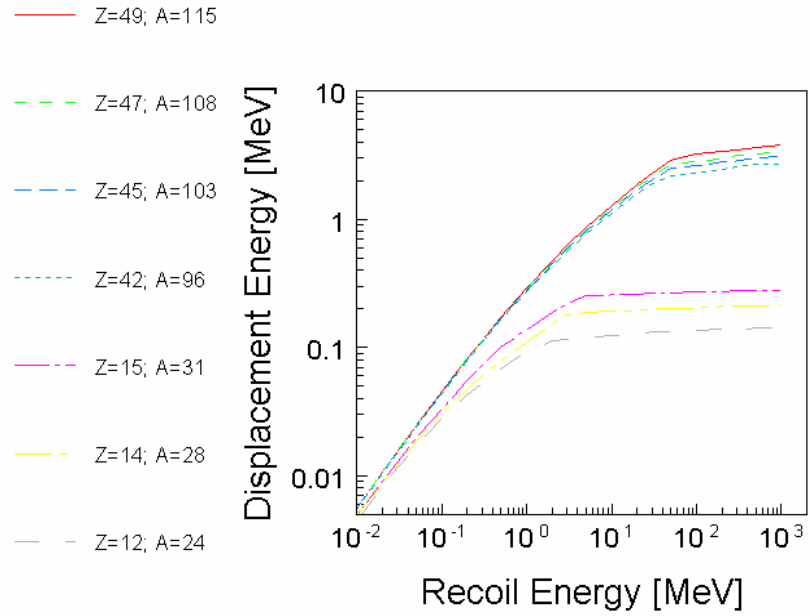

Figure 6. Lindhard factors in InP

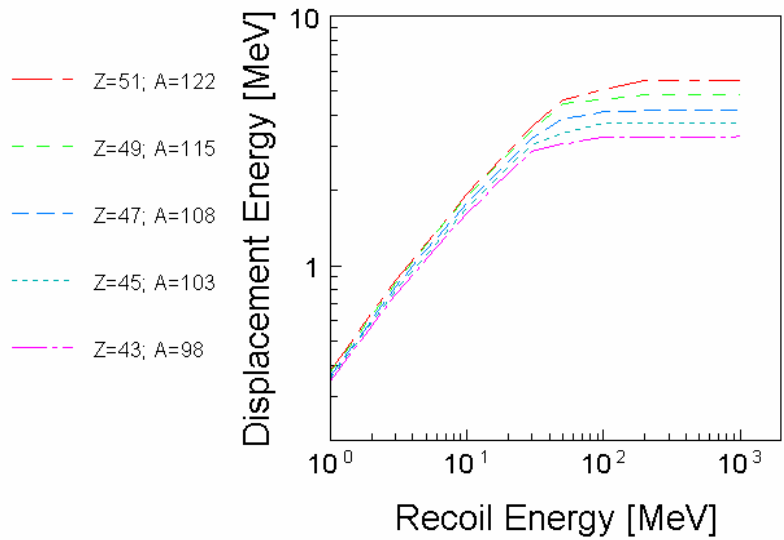

Figure 7. Lindhard factors in InSb

## II. Concentration of primary radiation induced defects

The concentration of the primary radiation induced defects per unit fluence in an (XY) compound has been calculated using the explicit formula:

$$CPD(E) = \frac{N_X}{2E_{d;X}} \int \sum_i \left(\frac{d\sigma}{d\Omega}\right)_{i;X} L(E_{Ri})_{XY} d\Omega + \frac{N_Y}{2E_{d;Y}} \int \sum_i \left(\frac{d\sigma}{d\Omega}\right)_{i;Y} L(E_{Ri})_{XY} d\Omega =$$

$$= \frac{1}{N_A} \left[ \frac{N_X A_X}{2E_{d;X}} (NIEL)_{XY}^{X\_family} + \frac{N_Y A_Y}{2E_{d;Y}} (NIEL)_{XY}^{Y\_family} \right]$$

where $E$ is the kinetic energy of the incident particle, $N_{X(Y)}$ is the atomic density of the $X(Y)$ element in the $XY$ compound, $A_{X(Y)}$ is the atomic number of the $X(or\ Y)$, $E_{d;X(Y)}$ - the threshold energy for displacements in the $X(Y)$ sublattice of $XY$, supposed independent on orientation, $E_{Ri}$ - the recoil energy of the residual nucleus produced in interaction $i$, $L(E_{Ri})$ - the Lindhard factor that describes the partition of the recoil energy between ionisation and displacements and $\left(d\sigma/d\Omega\right)_i$ - the differential cross section of the interaction between the incident particle and the nucleus of the lattice for the process or mechanism $(i)$, responsible in defect production. The atomic density $N_{X(Y)}$ in XY is a material parameter, which depends to the polytype through structural constants.

The formula gives also the relation with the non-ionising energy loss (NIEL). $N_A$ is Avogadro's number. It is important to observe that there exists a proportionality between the CPD and NIEL only for monoelement materials.

The CPD allows the comparison of the effects produced by the same or different particle in different materials, while NIEL is especially used for the comparison of the effects produced in the same material by different particles.

The values of the CPD produced by pions and protons in diamond, silicon, SiC and $A^3B^5$ compounds have been calculated and are presented in the following figures.

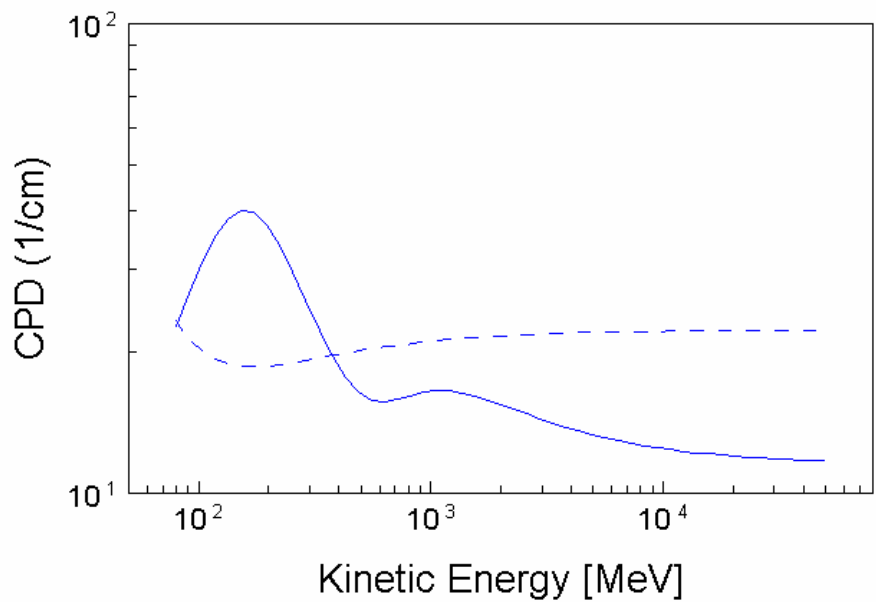

Figure 8. CPD induced by pions (continuous line) and protons (dashed line) in diamond (from [*I. Lazanu, S. Lazanu,, Nucl. Instr. Meth. Phys. Res. A  432, 374 (1999)*])

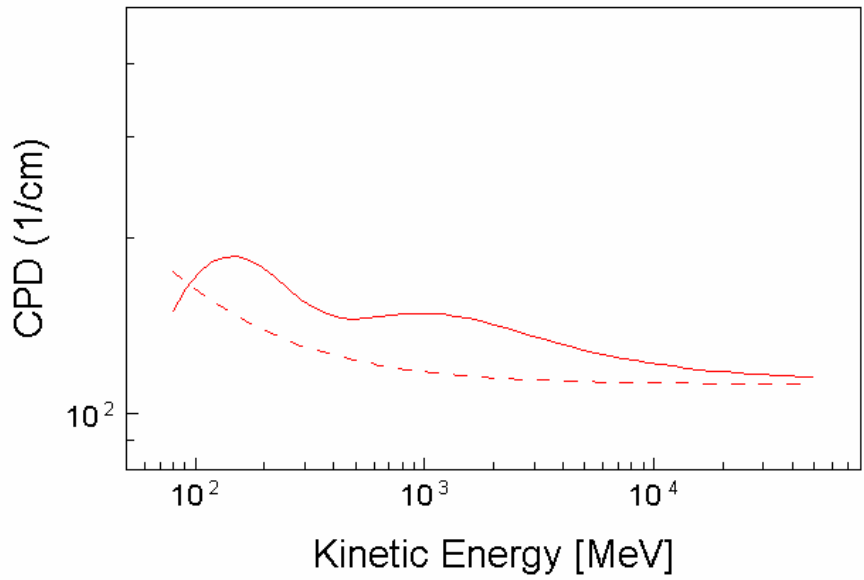

Figure 9: CPD induced by pions (continuous line) and protons (dashed line) in silicon (from [*S. Lazanu, I. Lazanu, Nucl. Instr. Meth. Phys. Res. A 419, 570 (1998)*], [*E. Burke et. Al., IEEE Transactions Nuclear Science, NS-33, 1276 (1986)*], [*A. Van Ginneken, preprint Fermilab, FN-522,1989*].)

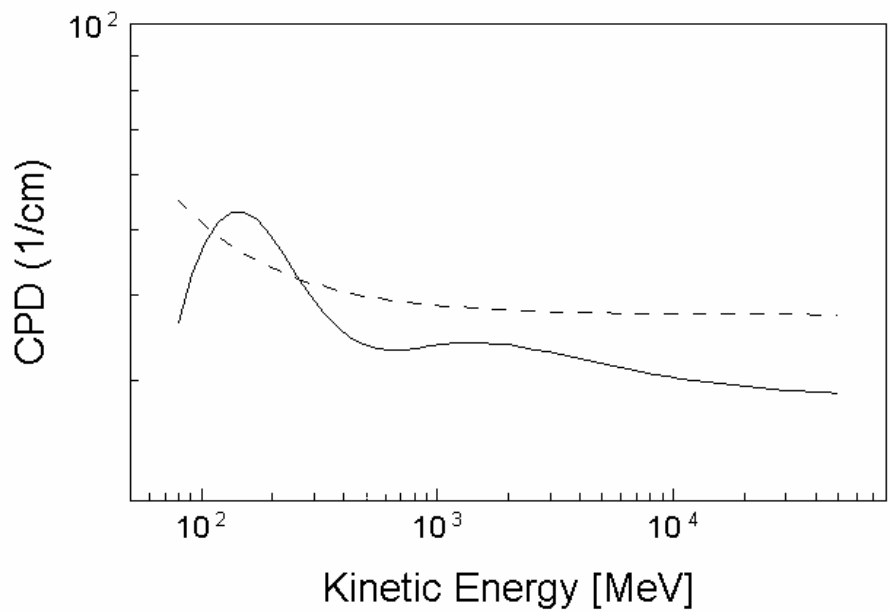

Figure 10: CPD induced by pions (continuous line) and protons (dashed line) in SiC (from [*S. Lazanu et al., Nucl. Instr. Meth. Phys. Res. A 485, 768 (2002)*])

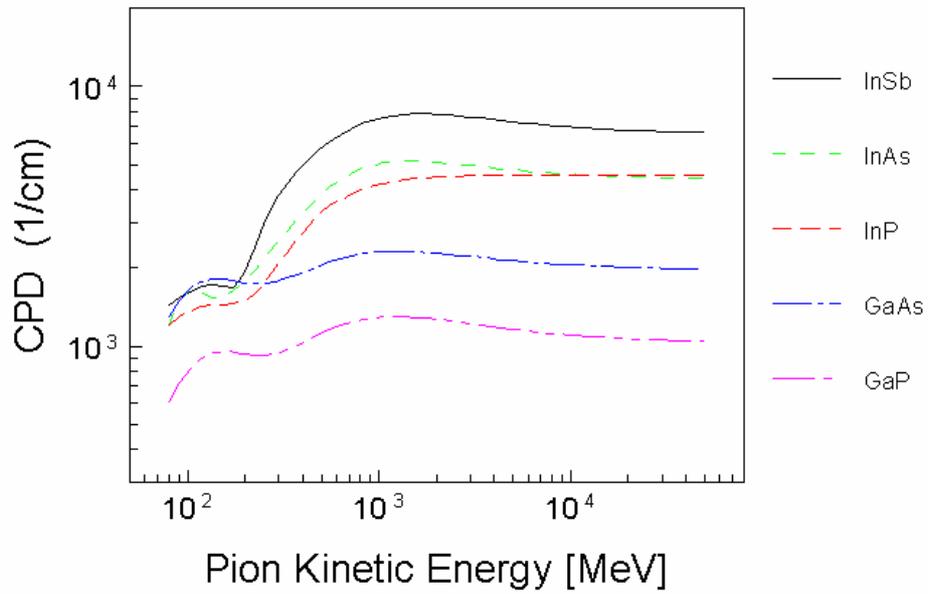

Figure 11: CPD induced by pions (continuous line) and protons (dashed line) in $A^3B^5$ compounds (from [*S. Lazanu et al., Nucl. Instr. Meth. Phys. Res. A 413, 242 (1998)*])

The difference between the positive and negative pions behaviour in diamond is indicated in figure 12.

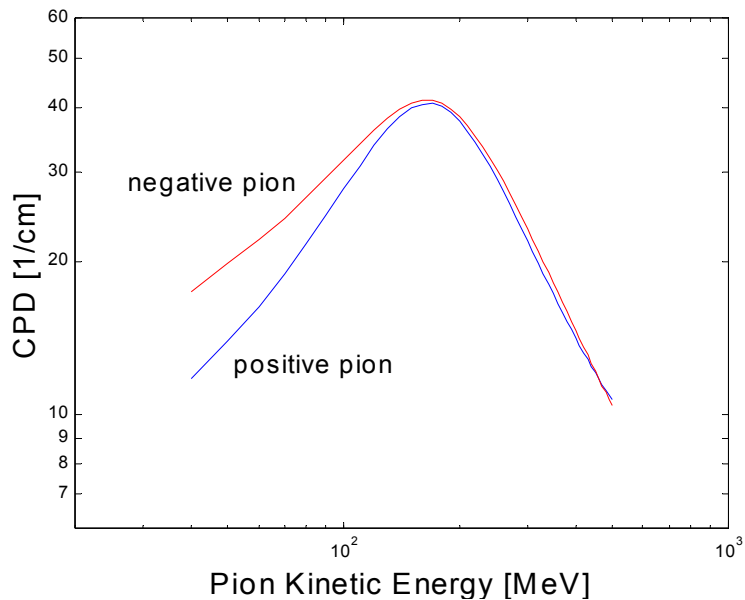

Figure 12: CPD versus kinetic energy for positive and negative pions in diamond (from [*I. Lazanu et al., Nucl. Instr. Meth. Phys. Res. A 406, 259 (1998)*]).